\documentstyle[aps,prl,multicol,amsmath,epsfig]{revtex}

\begin{document}

\newcommand{\prtl}{\partial}
\newcommand{\la}{\left\langle}
\newcommand{\ra}{\right\rangle}
\newcommand{\dla}{\la \! \! \! \la}
\newcommand{\dra}{\ra \! \! \! \ra}
\newcommand{\we}{\widetilde}
\def\Xint#1{\mathchoice
   {\XXint\displaystyle\textstyle{#1}}%
   {\XXint\textstyle\scriptstyle{#1}}%
   {\XXint\scriptstyle\scriptscriptstyle{#1}}%
   {\XXint\scriptscriptstyle\scriptscriptstyle{#1}}%
   \!\int}
\def\XXint#1#2#3{{\setbox0=\hbox{$#1{#2#3}{\int}$}
     \vcenter{\hbox{$#2#3$}}\kern-.5\wd0}}
\def\ddashint{\Xint=}
\def\dashint{\Xint-}

\draft

\title{The Origin of the Boson Peak and Thermal Conductivity Plateau 
in Low Temperature Glasses. }
\author{Vassiliy Lubchenko and Peter G. Wolynes}

\address{Department of Chemistry and Biochemistry, University of California 
at San Diego, La Jolla, CA 92093-0371}

\date{\today}

\maketitle

\begin{abstract}

We argue that the intrinsic glassy degrees of freedom in amorphous 
solids giving rise to the thermal conductivity plateau 
and the ``boson peak'' in the heat capacity at moderately low
temperatures are directly connected to those motions giving rise to
the two-level like excitations seen at still lower temperatures. 
These degrees of freedom can be thought of as strongly
anharmonic transitions between the local minima of the glassy 
energy landscape that are accompanied by ripplon-like domain wall 
motions of the glassy mosaic structure predicted to occur at $T_g$ 
by the random first order transition theory.
The energy spectrum of the vibrations of the mosaic depends on the 
glass transition temperature, the Debye frequency and the molecular
length scale.   
The resulting spectrum reproduces the experimental low temperature Boson
peak. The ``non-universality'' of the thermal conductivity plateau
depends on $k_B T_g/\hbar \omega_D$ and arises from calculable 
interactions with the phonons.
		
\end{abstract}


\begin{multicols}{2}
\narrowtext

A multitude of phenomena are observed in all low temperature glasses 
that can only be explained by the existence of excitations not 
present in crystals. At very low temperatures, the universal scalings
of heat capacity and thermal conductivity suggest these
excitations cannot be simply ascribed to the local molecular 
motions specific to each substance \cite{YuLeggett,FreemanAnderson}. 
Yet quantitative analysis of the experiments has generally relied 
on purely phenomenological models that invoke unrelated types
of excitations in different temperature regimes.
At liquid helium temperatures the excitations of glasses are well 
described as two-level systems \cite{AHV,Phillips,LowTProp}, 
but at somewhat higher 
temperatures the more mysterious ``boson peak'' appears in the heat 
capacity and the phonon mean free path falls precipitously, as if 
new scatterers became active, leading to a plateau in the thermal 
conductivity. Several different ideas about these additional 
excitations have been explored, ranging from harmonic excitations 
of a disordered lattice \cite{parisi} to anharmonic local modes of a 
``soft potential'' with distributed parameters \cite{soft}.

In contrast, here we suggest that like the two level systems \cite{LW}
these quantum excitations in glasses are intrinsic and essential 
relics of the non-equilibrium character of the glass after it is 
prepared. We show that quantized ``domain wall motions'' connected
with the mosaic structure of glasses predicted by the random first 
order transition theory of glasses \cite{PGW}, explain quantitatively the 
boson peak and conductivity plateau.
 
The random first order transition theory of the glass transition 
suggests there is a dynamical mosaic structure in classical 
supercooled  liquids \cite{XW}. This mosaic structure is directly 
manifested in the dynamical heterogeneity observed in supercooled 
liquids using single molecule experiments \cite{RusselIsraeloff} and nonlinear 
relaxation experiments \cite{Silescu}. Below $T_g$, the mosaic is spatially 
defined by the molecular motions that were not arrested at $T_g$, and is thus 
only dynamically detectable. We have shown 
these same motions at liquid helium temperatures would be quantized 
and can be described by the two state system phenomenology. This 
identification predicts the density of states arising from these 
motions and accounts for the previously unexplained universality of  
thermal conductivity in the liquid helium temperature regime. 
As the temperature is raised, however,  the theory indicates these 
excitations should
lose their strict two level character. The predicted temperature of 
the crossover to multilevel behavior coincides with the temperature 
regime where further excitable modes have been needed to explain
experiments. In this paper, we show explicitly how the number of 
extra excitable modes of the mosaic accounts quantitatively  
for the data, thus removing the need to invoke additional mechanisms,
although other contributions may well be present to some extent. 
The excitations in the boson
peak, like the two-level systems, active at lower temperatures, 
turn out to be collective motions of many particles ($\sim 200$) 
encompassing rearrangements of a typical domain frozen-in at the 
glass transition. The multilevel behavior of these domains can be 
pictured as involving the concommitant
excitation of the ``ripplon'' modes of the glassy mosaic along
with the transition between local minimum energy configurations
of a mosaic cell.

According to the random first order transition theory of 
supercooled liquids \cite{PGW}, metastable configurations begin to 
last locally longer than a few vibrations just below a dynamical
transition temperature $T_A$. 
Below this transition a glass forming liquid samples 
exponentially many metastable configurations much higher in energy 
than those of the corresponding crystal, but
which are long lived. Glassy dynamics is described by
the activated motions between minima.
The liquid can be visualized as a mosaic of domains which 
locally resemble low free energy configurations that are separated 
from each other by frustrated higher energy interfaces, 
or {\em domain walls}. 
The length scale of the mosaic and number density of these
walls is determined by the competition between the energy
cost of a wall and the entropic advantage of using the large
number of configurations. As temperature is lowered the 
configurational entropy per particle, $s_c$, decreases so that
the cooperative length grows. This leads to a larger activation
barrier that eventually gives relaxation times exceeding the 
laboratory time scale. The size and rate of motion of these
mosaic domains can be calculated without adjustable 
parameters once the liquid's heat capacity jump at the glass
transition is known \cite{XW}. In the classical regime this theory leads
to predictions of the average barrier and distributions of barriers
in quantitative agreement with experiment \cite{XWbeta}. 
According to this microscopic calculation, at the $T_g$ 
that corresponds with one hour relaxation time, the size of the 
cooperative region $\xi$ is expected to vary only a little from 
substance to substance. A mosaic cell allows for at 
least one alternative kinetically accessible configuration at $T_g$.
These alternate states therefore have spectral density 
$\sim 1/T_g$  per domain of size $\xi^3$ \cite{LW}. This agrees
with the measured density of two-level systems \cite{LowTProp}.
While activated at $T_g$, below $T_g$, motion between these 
alternative states remains possible, switching to coherent resonant 
tunneling at a temperature proportional but smaller than the Debye 
temperature: $\sim (a/\xi) T_D/2\pi$ \cite{LW}.

\begin{figure}[tbh]
\centerline{\epsfig{file=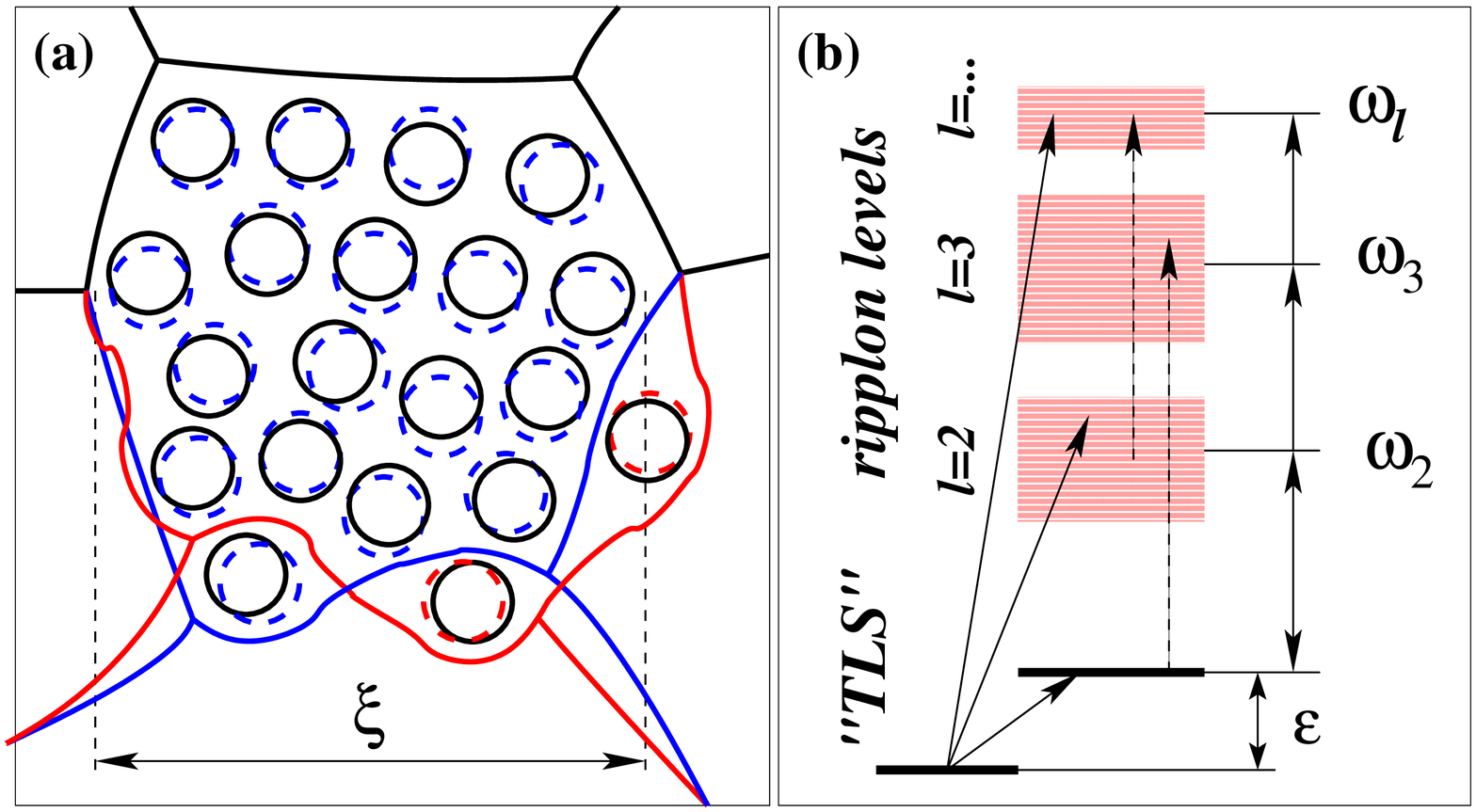, height=4.5cm}}
\caption{(a) Schematic of a tunneling mosaic cell is given, with 
doubled circles denoting atomic tunneling displacements. 
The boundary's location is variable, as illustrated 
(in an exagerated manner) by the blue and red lines. 
The domain wall distortion amounts to populating  
ripplon states on top of the structural transition energy $\epsilon$, 
as shown in (b). All transitions exemplified
by solid lines involve tunneling between the intrinsic states 
and are coupled linearly to the lattice distortion and contribute 
the strongest to the phonon scattering. The ``vertical''
transitions, denoted by the dashed line, are coupled to the higher order
strain; they contribute only to the Rayleigh scattering 
(which is too weak to account for the plateau \protect\cite{ACAnd_Phi}).}
\label{en_levels}
\end{figure}

At low temperatures the two-level system excitations involve
tunneling of the mosaic cells typically containing $N^* \simeq 200$ 
atoms. The tunneling path involves stagewise motion of the wall separating 
the distinct alternative configurations through the cell
untill a near resonant state is found. At higher temperatures,
other final states are possible since the exact number and identity 
of the atoms that tunnel can vary (see Fig.\ref{en_levels}a). 
These new configurations typically will be like the near resonant 
level but will also move a few atoms at the boundary, i.e. at the 
interface to another domain.
These fluctuations of the domain boundary shape can be visualized
as domain wall surface modes (``ripplons''). They cost a surface
energy that varies with the size of the domain and  have kinetic
energy consistent with the mass of the moving domain wall.
It is not surprising that the ripplon's frequencies turn out 
to be proportional to $\omega_D$, the basic quantum energy scale in the 
system just as does the crossover temperature. A detailed calculation
of the ripplon spectrum would require a considerable knowledge of the
topology and statistics of the mosaic. At each temperature below $T_A$ 
the domain wall foam is an equilibrium structure made up of nearly
flat patches. According to the random first order transition theory, 
the effective surface tension depends on the curvature and vanishes
at large radius of curvature as $\sigma(r) \propto r^{-1/2}$. 
To approximate the spectrum we notice that the ripples of wave-length larger 
than the size of a patch will typically sense a roughly spherical surface of 
radius $R = \xi (3/4\pi)^{1/3}$. The surface tension of the mosaic
has been calculated from the classical microscopic theory and is given
by $\sigma(R) = \frac{3}{4}
(k_B T_g/a^2) \, \log((a/d_L)^2/\pi e) (a/R)^{1/2}$ \cite{XW}, where
$d_L/a$ is the universal Lindemann ratio. By itself such a tension
would collapse the mosaic but this tension should be compensated
by stretching the frozen-in outside walls of neighboring mosaic cells. 
This compensating effect 
can be  approximated by an isotropic positive pressure of a ghost 
(i.e. vanishing density) gas on the inside. Calculating the 
frequencies of the surface eigen-modes of a hollow sphere, subject only
surface tension, is a classic problem of mathematical physics
\cite{Morse}. Accounting for the unusual $r$ dependence of
the surface tension $\sigma(r) \propto r^{-1/2}$ modifies the standard 
result for the frequencies is by a factor of $9/8$. The 
eigenmodes have frequencies 
$\omega_l^2 = \frac{9}{8} (\sigma/\we{m}_W R^2) (l-1)(l+2)$, 
where $\we{m}_W = (d_L/a)^2 \rho a$ is the domain 
wall mass per unit area \cite{LW} and $\rho$ is the mass density
of the glass. 
The $l$-th mode of a sphere is $(2 l+ 1)$-fold degenerate. Using 
$k_B T_g \simeq \rho c_s^2 a^3 (d_L/a)^2$ \cite{LW}, one finds then
$\omega_{l} \simeq 1.34 \, \omega_D (a/\xi)^{5/4} \sqrt{(l-1)(l+2)/4}
\simeq  0.15 \, \omega_D \, \sqrt{(l-1)(l+2)/4}$.
Because of the universality of the $(a/\xi)$ ratio \cite{XW}, 
$\omega_l$ is a multiple of the Debye frequency. Due to the 
material's discreteness, there are no harmonics of higher than 
$\left(\frac{3}{4 \pi} N^* \right)^{1/3} [(R-a/2)/R] \simeq 9..10$th order, 
a relatively large number, which makes the continuum 
approximation plausible. The lowest allowed ripplon mode is $l=2$, 
and has a frequency of $\sim 1$THz for silica. This is in agreement
with the Boson peak in the inelastic neutron scatering data 
\cite{bos_peak}). $l=1$ corresponds to a domain translation and is
accounted for by the underlying phonon momentum non-conserving 
tunneling transition itself. The $l=0$ is a uniform domain dilation 
relevant to aging below  $T_g$. The existence of the domain 
wall vibrations lets us visualize the 
multilevel character of the tunneling centers as exhibited at 
temperatures above the TLS regime. A schematic of the resultant 
droplet quantized energy levels is shown in Fig.\ref{en_levels}b.

\begin{figure}[htb]
\centerline{\epsfig{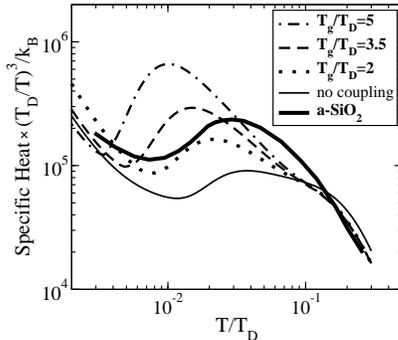}}
\caption{The heat capacity per domain, as follows from the derived 
TLS $+$ ripplon density of states, 
divided by $T^3$. This includes the Debye contribution.
The thin line neglects phonon coupling and has $T_g/T_D=4$.
The experimental curve for 
amorphous silica from \protect\cite{Pohl}, originally given in J/gK$^4$, 
is shown multiplied by $\hbar^3 \rho c_s^3 (6 \pi^2) (\xi/a)^3/k_B^4$, 
where we used $\omega_D = (c_s/a)(6\pi^2)^{1/3}$, $(\xi/a)^3=200$
\protect\cite{LW}, $\rho = 2.2 \mbox{g/cm}^3, c_s=4100 \mbox{m/sec}$  
and $T_D = 342 K$ \protect\cite{FreemanAnderson}.
Using the appropriate  value of $k_B T_g/\hbar \omega_D = 4.4$
for a-SiO$_2$ would place the peak somewhat lower in temperature than 
observed.} 
\label{bump}
\end{figure}

The (classical) density of energy minima of a domain is $n(\epsilon)
=\frac{1}{k_B T_g} e^{\epsilon/k_B T_g}$ \cite{LW}, where the energy zero
corresponds to the (high-energy) configuration frozen-in at $T_g$. 
The negative $\epsilon$'s correspond 
to some of the very numerous but mostly unavailable lower lying energy 
states, accessed by tunneling. If we refer to the lower state of the
resonant pair as the local ground state, the excitation energy 
density becomes $n(\epsilon) =\frac{1}{k_B T_g} e^{-|\epsilon|/k_B T_g}$ 
($\epsilon > 0$). The partition function of a domain with
an excitation energy
$\epsilon$, possibly accompanied by populating vibrational
states on top, is given by $Z_{\epsilon}=1+e^{-\beta \epsilon}
\prod_l Z_l^{2l+1}$, where $Z_l \equiv 1/(1-e^{-\beta \hbar \omega_l})$
is the partition function of an $l$th order ripplon mode. 
Apart from the excitation $\epsilon$, we assume here 
each ripplon has a nearly harmonic spectrum. Note, our energy level scheme 
automatically insures that the atomic motions near the thermally 
inactive defects will only contribute to the regular lattice 
vibrations. The specific heat per domain, calculated 
from  $Z_\epsilon$ and averaged over $n(\epsilon)$, is
shown in Fig.\ref{bump}. 

In order to estimate the scattering by the domain tunneling accompanied
by the ripplons and thus the heat conductivity, we need to know the 
effective {\em scattering} density
of states, the transition amplitudes and the coupling to the phonons.  
Any transition in the domain accompanied by a change in its internal 
state is coupled to the gradient of the elastic field with 
energy $g \sim \rho c_s^2 \int d {\bf s} \cdot {\bf d} ({\bf r})$,
where ${\bf d}({\bf r})$ is the molecular displacement at the droplet 
edge due to the transition \cite{LW}. 
Therefore any transitions between groups
marked with solid lines in Fig.\ref{en_levels}b are coupled to the phonons 
with the same strength as the underlying (TLS-like) transition,
shown to be $g \simeq \sqrt{k_B T_g \rho c_s^2 a^3}$ \cite{LW}, 
and no selection rules apply for the change in the ripplon
quantum numbers because of the strong anharmonicity.  
We do not possess detailed information on the transition amplitudes,
however they should be on the order of the transition frequencies themselves,
just as for those TLS's that are primarily responsible 
for the phonon absorption at the lower $T$ whose transition 
amplitudes are also comparable to the total energy splitting. 

Since all transitions couple equally to the phonons, 
we can now calculate the density of the scattering states.
If $\epsilon > 0$, the phonon absorbing transition occurs from 
the ground state. The corresponding total number of ways to absorb
a phonon of energy $\hbar \omega$
is $\rho (\omega) = \int_0^{\infty} d\epsilon \, 
n(\epsilon) \sum_{\{n_{lm}\}} 
\delta(\hbar \omega-[\epsilon+\sum_{lm} n_{lm} \hbar \omega_{lm}]) = 1/k_B T_g 
\sum_{\{n_{lm}\}}  \theta(\omega-\sum_{lm} n_{lm} \omega_{lm}) 
e^{-\beta_g \hbar (\omega-\sum_{lm} n_{lm} \omega_{lm})}$,
where we sum over all occupation numbers of the ripplons with
quantum numbers $l,m$ ($m=-l..l$). 
The transition may also occur from the higher energy conformational
state ($\epsilon <0$ in Fig.\ref{en_levels}b), and we have 
to compute $\rho(\omega)$ from these states too. To find $\rho(\omega)$, 
we compute the cumulative density of states 
$N_E (\omega) \equiv 
\int_{-\infty}^0 d\epsilon \, \rho (\omega; -\epsilon<E) = \int_0^{E} 
d\epsilon \, n(\epsilon) \sum_{\{n_{lm}\}} 
\delta(\hbar \omega-[\sum_{lm} n_{lm} \hbar \omega_{lm}-\epsilon])$.
In a calculation parallel to obtaining  $Z_l$, 
this can be evaluated using an integral representation
of the step function $\theta$ along with the use of 
steepest descent.
The scattering from excited states is proportional to
$\rho_{exc}(\omega,T) 
\equiv \int_0^{\infty} 
d E \, f(E,T) \, \prtl N_E (\omega)/\prtl E$, where 
$f(E,T) \equiv 2/(e^{\beta \epsilon}+1)$ gives the appropriate 
Boltzmann weights.

Accurate calculation of the heat conductivity requires solving
a kinetic equation for the phonons coupled with the multilevel 
systems, which would account for saturation effects etc. We utilize
instead a single relaxation time approximation for each phonon
frequency. The Fermi golden rule yields for the scattering rate 
of a phonon with $\hbar \omega \sim k_B T$ the relation 
$\tau_{\omega}^{-1} \sim   \, \omega \frac{\pi g^2}
{\rho c_s^2} \, [\rho(\omega) + \rho_{exc}(\omega,T)]$.
The heat conductivity then equals $\kappa = 
\frac{1}{3} \sum_\omega l_{mfp}(\omega) C_\omega c_s$. The mean free 
path cannot be less than the phonon's wave-length $\lambda$ 
(which occurs at the Ioffe-Riegel condition). We account for multiple 
scattering effects by putting $l_{mfp} = c_s \tau_{\omega} + \lambda$.
At high $T$, the heat is not carried by ``ballistic'' phonons, 
but rather is transfered by a random walk from site to site, 
as originally anticipated by Einstein \cite{Einstein} for homogeneous 
isotropic solids.
 
The coupling of the multilevel excitations to phonons leads
to significant frequency shifts and damping of the resonant 
transitions. To compute these coupling effects, we replace
the discrete summation over the different uncoupled 
harmonics $\sum_l \int d\omega \, \delta(\omega-\omega_l)$ by 
integration over ``lorentzian'' profiles $\sum_l \int d\omega \, 
\frac{\gamma_\omega/\pi}{[\omega-\omega_l(\omega)]^2+\gamma_\omega^2}$, 
where $\gamma_\omega$ is a (frequency dependent) friction coefficient
and $\omega_l(\omega)$ is the renormalized ripplon frequency,
which has been shifted due to the corresponding dispersion 
effects.  The total relaxation rate of a transition involving more than
one \begin{figure}[tbh]
\centerline{\epsfig{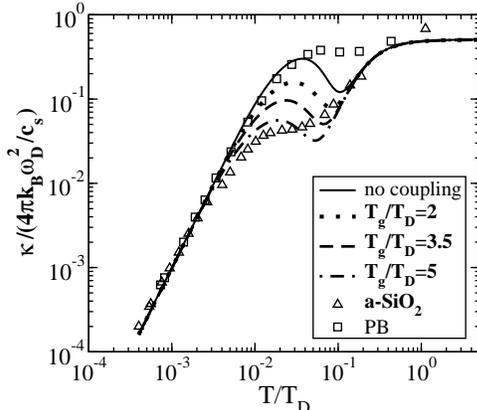}}
\caption{The predicted low $T$ heat conductivity.
The ``no coupling'' case neglects phonon coupling effects on the 
ripplon spectrum. The (scaled) experimental data are taken from 
\protect\cite{smith_thesis} for a-Si ($k_B T_g/\hbar \omega_D \simeq 4.4$) and 
\protect\cite{FreemanAnderson} for polybutadiene 
($k_B T_g/\hbar \omega_D \simeq 2.5$). The empirical universal lower $T$ 
ratio $l_{mfp}/l \simeq 150$ \protect\cite{FreemanAnderson}, used 
explicitly here to superimpose our results on the experiment, was predicted 
by the present theory earlier within a factor of order unity 
\protect\cite{LW}.}  
\label{plateau}
\end{figure} \vspace{-.2cm} mode is thus the sum of the 
inverse life-times
of the participating modes. This would be entirely correct 
for a frequency independent $\gamma$, but should be still an adequate 
approximation at the low $T$ end of the plateau, where the absorption 
is mostly due to single ripplon mode processes. The mode decay rate 
$\gamma_\omega$ is the phonon irradiation induced drag equal 
in the lowest order of perturbation to
$\gamma_\omega = \frac{g^2}{4 \pi \rho c_s^2} (\omega/c_s)^3
\simeq \frac{3 \pi}{2 \hbar} k_B T_g (\omega/\omega_D)^3$, where we do not 
distinguish between the longitudinal and transverse phonons.
The corresponding (frequency dependent) ripplon frequency shift
required by Kramers-Kronig relation gives a renormalized frequency 
$\omega_l(\omega) = \omega_l - \frac{3}{2 \hbar} k_B T_g
(\omega_c/\omega_D)^3 \dashint_0^{\omega_c} 
\frac{d \omega' (\omega'/\omega_c)^3}{\omega'-\omega}$, where
the principal value integral (numerically $\sim 1$) is a slow 
(but sign-changing!) function of $\omega$
and the cut-off frequency $\omega_c$.  $\omega_c$ is
of the order (but greater than) $(a/\xi) \omega_D$, since the phonons 
with wave-length shorter than $\xi$ cause an 
increasingly smaller effective gradient of the phonon field  as sensed 
by a region of size $\xi$. 
Although our approximation for $\gamma_\omega$
and $\omega_l (\omega)$ should break down in detail for 
$\omega \gtrsim \omega_c$,
it makes little difference computationally, as the damping is
already very intense at this point. Both the frequency shift and the
damping broaden the absorption peak, the former being quantitatively
more important. While the peaks' positions are 
determined by the quantum energy scale $\omega_D$, the 
spectral {\em shifts} are proportional to $T_g$. 
The non-universality of the $k_B T_g/\hbar \omega_D$ ratio, which varies 
over the range of $2$ to $5$, thus
yields a non-universal position of the plateau.
This is consistent with the experimentally observed material 
dependence in this regime. 
In Fig.\ref{plateau}, we show the resultant heat conductivities 
for different values of the ratio $k_B T_g/\hbar \omega_D$, 
using the specific cutoff $\omega_c = 1.8 (a/\xi) \omega_D$. 
The non-monotonic behavior exhibited by this approximation
in the plateau region is likely an artifact of neglecting thermal
saturation in the kinetic theory. Assuming heat transport is dominated
by thermal phonons suggests $\rho_{exc} (\omega,T)
\rightarrow \rho_{exc} (\omega, \sim \hbar \omega/k_B)$.
This leads to a temperature independent $l_{mfp} (\omega)$ and 
a strictly horizontal plateau.

The coupling effects on the heat capacity can be obtained
by replacing each $Z_l$ with the formula for a damped oscillator.
The resulting bump in $c/T^3$, shown in Fig.\ref{bump}, is therefore
also non-universal depending on $k_B T_g/\hbar \omega_D$, being
smaller for organic glasses, as seen in experiment. 

This work was supported by NSF grant CHE-9530680.

\end{multicols}


\begin{references}   

\vspace{-1cm}

\bibitem{YuLeggett} C.C.Yu and A.J.Leggett, Comments Cond. Mat. Phys.,
{\bf 14}, 231 (1988); A.J.Leggett, Physica B {\bf 169}, 322 (1991).

\bibitem{FreemanAnderson} J.J.Freeman and A.C.Anderson,  Phys. Rev. 
{\bf  B 34}, 5684 (1986).

\bibitem{AHV} P.W.Anderson, B.I.Halperin and C.M.Varma,
Philos. Mag. {\bf 25}, 1 (1972).

\bibitem{Phillips} W.A.Phillips, J. Low Temp. Phys. {\bf 7}, 351
(1972).

\bibitem{LowTProp} Amorphous Solids: Low-Temperature Properties,
edited by W.A.Phillips, Springer-Verlag Berlin Heidelberg New York, 
1981.

\bibitem{parisi} T.S.Grigera, V.Mart\'{i}n-Mayor, G.Parisi, and
P.Verrocchio, Phys. Rev. Lett. {\bf 87}, 085502 (2001).

\bibitem{soft} Yu.M.Galperin, V.G.Karpov, and V.I.Kozub, 
Adv. Phys. {\bf 38}, 669 (1991).

\bibitem{LW} V.Lubchenko and P.G.Wolynes, Phys. Rev. Lett.
{\bf 87}, 195901 (2001); cond-mat/0105307.

\bibitem{PGW} P.G.Wolynes, Acc. Chem. Res. {\bf 25}, 513 (1992);
T.R.Kirkpatrick, D.Thirumalai, Transp. Theor. Stat.Phys. {\bf 24}, 
927 (1995) and references therein.

\bibitem{XW} Xiaoyu Xia and P.G.Wolynes, PNAS {\bf 97}, 2990 (2000);
cond-mat/9912442.

\bibitem{RusselIsraeloff} E.V.Russel and N.E.Israeloff, Nature
{\bf 408}, 695 (2000).

\bibitem{Silescu} H.Silescu, J.Non-Cryst. Solids {\bf 243}, 81 (1999).

\bibitem{XWbeta} Xiaoyu Xia and P.G.Wolynes, Phys. Rev. Lett.
{\bf 86}, 5526 (2001); cond-mat/0008432.

\bibitem{ACAnd_Phi} A.C.Anderson in \cite{LowTProp} and references
therein.

\bibitem{Morse} P.M.Morse and H.Feshbach, ``Methods of Theoretical
Physics'', McGraw-Hill, 1953; v.2 p.1469.

\bibitem{bos_peak} A.Wischnewski, U.Buchenau, A.J.Dianoux, 
W.A.Kamitakahara, J.L.Zaretsky, Phys. Rev. B {\bf 57}, 2663 (1998).

\bibitem{Pohl} R.O.Pohl in \cite{LowTProp}.

\bibitem{Einstein} A.Einstein, Ann. Phys. {\bf 35}, 679 (1911). 

\bibitem{smith_thesis} T.L.Smith, Ph.D. thesis, University of Illinois, 1974. 


\end{references}
\end{document}